# A.V. Usova's Contribution to the Field of Concept Learning in Physics Classroom

**Oleg Yavoruk**
Yugra State University
Department of Physics and Technical Disciplines
Russian Federation

**Abstract**

A.V.Usova (1921-2014) has always been one of the leading figures in Russian physics education. Her theory of physics concept formation was formulated during the 1970s and the 1980s and directly influenced the process of physics education in the 20th and the 21st century. Over the years there have been a lot of theories of concept formation. Her work contributed to our understanding of concept formation (learning, teaching) and the contemporary physics learning process. She formulated her original views on the problem of concept formation independently of Western researchers. She is perhaps the most important Russian educational theorist in the field of concept learning. A.V.Usova suggested to physics teachers the model of concept formation that describes: methods of learning concept in physics classroom, conditions of successful concept formation in physics teaching; structure of complex scientific concepts formation (stages of concept formation), the influence of interdisciplinary teaching on the scientific concept formation, criteria and levels of physics concept formation, methods and techniques of analysis of the quality of concept formation, the role of educational observation and experimentation in the scientific concepts formation, methodology formation for complex physics concepts "work" and "energy". This paper deals with both the historical A.V.Usova's contribution review and also issues raised by post-Usova approaches.

*Oleg Yavoruk*

## Introduction

Antonina Vasilyevna Usova was born on August 4, 1921 in Bashkiria, USSR. In 1946 she graduated from Kazan State University. She worked at Chelyabinsk State Pedagogical University in the Physics Teaching Department in the period 1951-2014. From 1973 to 2006 she is Head of the Department. She was a famous Russian scientist who made a significant contribution not only to the teaching of physics, but also to general didactics and pedagogy. She earned her Doctorate degree in Pedagogical Sciences in 1970, and became professor in 1973. She died on August 8, 2014, in Chelyabinsk (South Urals of Russia).

She dealt with the problem of concept formation since 1965 (Dammer & Krestnikov, 2011). At that time, physics was one of the most important disciplines in Soviet secondary public schools.

International scientific communication was very difficult due to the Iron Curtain (1945-1991) that symbolized the conflict, ideological and physical boundaries between so-called communist and capitalist states. Nevertheless, some works of prominent American psychologists (e.g. Jerome Bruner) have been translated into Russian. In addition, there were strong academic psychological scholars in the Soviet Union (e.g. Lev Vygotsky). We find references to them in the major works by A.V.Usova. However, most of the works of American psychologists and pedagogues were unknown behind the Iron Curtain, as well as many works of Soviet psychologists, pedagogues and educators remain unknown to Western science. So a lot of problems have been solved independently, although sometimes in a similar way.

Accurate translation from Russian into English is quite difficult. Many tones of Russian scientific language disappear when presenting the same material in English. Probably, this problem can be solved by a method of multivariate representation, when the same thing is spoken in different ways and with different words over and over again. This is very important because we are talking about the life of a particular person, who played a significant role in the educational system of my country (Belozyortsev, 2006).

I am so grateful to people who told me about A.V.Usova's life, and the responsibility for any omissions, errors, discrepancies of this paper, of course, is mine.

## The System of Physical Knowledge and Generalized Plans of Learning

One of the major problems that must be solved by schools is acquiring the scientific knowledge system. Through logical and genetic analysis professor Usova identifies the following elements of knowledge: scientific facts, concepts, laws, theories, etc. Considering them, she indicates their universality and interrelation. A simple classification of physical concepts implies a division into entities, phenomena, properties and quantities.

Her pedagogical discoveries are the generalized plans (i.e. algorithms, strategies, guides, sequences, schemata, rules, techniques) of studying or executing (for concepts, laws, theories, methods, instruments, etc.).

How to study physical phenomena? How to study physical laws? How to study physics entities? Properties? Quantities? What is the structure of the element of knowledge? What do you need to know about this element of knowledge? When can I say about myself that I do know the law? The concept? How to describe a physical phenomenon? Property? Quantity? Physical law, theory? What should I learn if I want to study this theory alone (independently, by self-study)? How to study physics? How to learn physics? How to teach physics?

      Complete list of generalized plans includes also rules for conducting observations, experiments, and apparatus learning. These short generalized plans are of universal significance. A unifying framework for concept-learning plans is of special importance for A.V.Usova.

**Physics Phenomena**
1. External features of the phenomenon.
2. Conditions of the phenomenon.
3. The essence of the phenomenon, its mechanism.
4. Relation of the phenomenon with other phenomena.
5. Quantitative characteristics of the phenomenon.
6. Application of the phenomenon.
7. Prevention of harmful effects of the phenomenon.

**Physics Quantities**
1. What phenomenon or property is characterized by this quantity.
2. Definition of the physics quantity.
3. Specific characteristics of this quantity (scalar or vector, basic or derivative)
4. Definitional formulas for this quantity.
5. Units of this quantity according SI (International System of Units).
6. Methods of measuring.

**Laws of Physics**
1. What concepts are related to the law of physics?
2. Formulations of the law.
3. The mathematical expressions of the law.
4. Limitations of the law.
5. Who discovered the law? When and how?
6. Experimental evidence of the law.
7. Examples of natural phenomena, where the law acts.
8. Practical application of the law.

**Theories of Physics**
1. Scientific facts on which the theory is based.
2. Key concepts of the theory.
3. The basic assertions, postulates, laws, principles of the theory.
4. Mathematical formalism of the theory, fundamental equations.
5. Results, findings, and predictions of the theory.
6. Limitations of the theory.

**Plan of Observations**
1. Time of the observation and information about the observer.
2. Purpose of the observation (it is formulated by a teacher or chosen by the students).
3. Description of the observed object.
4. Equipment of the observation (it may be selected by students).
5. Conditions of the observation.
6. Description of the observational procedure.

7. Results of the observation.
8. Analysis of results and conclusion from them.

**Concept Formation**

A.V.Usova's book "The Formation of Scientific Concepts among Students in the Learning Process" which was published in Russian, detailed the psychological and didactic basis for the formation of physics concepts. The first edition of this book was published by the Publishing House "Pedagogy" in 1986. Publishing House "University of RAO" released the second edition of this book in 2007.

The book has played the key role not only in the development of methods of physics teaching; it is a significant contribution to the modern didactics, so we consider it in detail.

**The Main Topics of the Book**
1. Methods of learning concepts in physics classroom.
2. Conditions of successful concept formation in physics teaching.
3. Structure of complex scientific concepts formation (stages of concept formation).
4. The influence of interdisciplinary teaching on the scientific concept formation.
5. Criteria and levels of physics concept formation.
6. Methods and techniques of the quality analysis for concept formation.
7. The role of educational observation and experimentation in the scientific concepts formation.
8. Methodology of formation for complex physical concepts "work" and "energy".

In accordance with the views of A.V.Usova, concept is the knowledge of the essential properties (sides) of objects and phenomena of reality, knowledge of significant connections and relationships between them. Other definitions of concept are not rejected; they are presented in her book and analysed in detail.

In the first chapter of the book "The concept as a logical category", A.V.Usova considers concept as a logical and epistemological category: she gives a rigorous scientific definition of it; describes the connections and relationships between concepts; and discusses methods on acquaintance to the concepts in the case when the definition is not possible or is not required.

In the second chapter, "The development of concepts in the scientific and academic knowledge" A.V.Usova makes a consistent presentation of the psycho-physiological foundations of concept formation, clarifies their nature and role in scientific cognition, features the development of concepts in science, shows the importance of the formation of scientific concepts in the learning process, describes the role of prescientific ideas in it, reveals the essence of the process of concept acquisition, and lists criteria and levels, typical mistakes and difficulties.

The third chapter deals with the ways of scientific concepts formation in the learning process. A.V.Usova analyses the different points of view on the process of mastering scientific concepts by students, different techniques of concept formation used in school practices, and lists the necessary elements and stages of the formation of complex physics concepts.

Concretization and generalization of concepts are represented in the fourth chapter of the book (both theoretical and practical issues).

The fifth chapter is devoted to investigating conditions which facilitate concept acquisition; describing the quality evaluation of it; characterizing opportunities of interdisciplinary relationships, educational observation and experimentation in concept formation.

The sixth chapter is devoted to methodology of the fundamental physics concepts formation (such as "work" and "energy").

The problem of concept formation was derived from problems of the science foundation learning, but concepts of knowledge play a particular role. Scientific concepts are one of the most important components of the knowledge system. The emergence of concepts is the result of the scientific cognition process of the world (Usova, 1986). Many well-known researchers have paid attention to the problem of formation of concepts (Vygotsky, 1962; Bruner, 1960). The peak of research on the formation of concepts accounted for the period 1960-1980. However, progress in solving the problem has not resulted to significant advances in educational practice. This requires a broad discussion between teachers.

**Concept Formation Conferences**

Conferences on the formation of concepts have been organized in the Soviet Union and then in the Russian Federation since 1971 (Chelyabinsk, South Urals). The Mastermind of these conferences was A.V.Usova. She organized them annually in the same month- every May.

Participants of these conferences debated the problems of the established concepts formation, interpretation of new scientific concepts necessitated by the development of science. For more than a third of a century, participants studied a huge number of possibilities of concept formation. The emphasis has been on the methods of teaching, which would be found from the viewpoint of concept formation. The problem of concept formation was discussed not only by philosophers, psychologists and educators, but also by physicists, astronomers, mathematicians, chemists, biologists, ecologists, philologists, sociologists, etc. (Usova, 1994).

Proceedings of conceptual conferences discussed by researchers have been the basis for hundreds of master's, and doctoral dissertations, for thousands of methodological recommendations, teaching instructions, textbooks and manuals, scientific articles and monographs (Krestnikov, 2006).

**Other Areas of Educational Research**

A.V.Usova is perhaps the most important educational theorist in physics teaching of Russian education. She made important contributions to the development of the fields of pedagogical science concerning learning, teaching, studying, etc. Books of A.V.Usova were used in the preparation of thousands of school physics teachers (Usova, 1988; 2002), and not only in the Soviet Union and Russia. Her book "Methodology of Physics Teaching" was translated into Spanish: "Metodología de la Enseñanza de la Física" (Orejov & Usova, 1980).

**Problems of Education Considered by A.V.Usova**
1. Interdisciplinary relationships in the learning process.
2. Organization of self-learning of students.
3. General and specific issues of physics teaching at schools and universities.
4. Formation of general learning skills.
5. Teaching students to solving of physics problems.
6. Methodology of pedagogical experiment (elemental and operational analysis).
7. Issues of polytechnic education.
8. Theory and practice of developing education.

**Conclusion**

Psychological and didactical aspects of the scientific concepts formation are important to use in the preparation of textbooks and teaching manuals and instructions in physics education. Didactic theory presented by A.V.Usova allows one to find the optimal structure for the learning of physics.

A.V.Usova clarified the essence of the concept formation; analyzed the different views on the concepts learning and methods for their formation, explored the conditions for successful formation of concepts, and ways of improving their quality. She paid special attention to the analysis of errors in the acquiring of concepts, as well as their causes and prevention.

Concept formation theories and their roles in contemporary physics teaching have been theorized over the last 100 years, and many interesting decisions have been found by A.V.Usova. Successful solution of this problem is seen by the teachers mastering the theoretical foundations of the concept formation.

Over this long period of time, both in society and in the education system enormous changes have taken place, however the theory proposed by A.V.Usova remains relevant today.